\documentclass[twocolumn, prl]{revtex4}
\usepackage{graphicx}
\usepackage{epsfig}
\usepackage{color} %
\usepackage{soul}
\begin{document}
\title{Potential Scattering on a Spherical Surface} 

\author{Jian Zhang$^{\ast}$ and Tin-Lun Ho$^{\dagger, \dagger\dagger}$}
\affiliation{$^{\ast}$ Institute of Nuclear  and New Energy Technology,  
 Collaborative Innovation Center of \\
Advanced Nuclear Energy Technology,  Key Laboratory of Advanced Reactor Engineering and Safety of Ministry of Education, 
Tsinghua University, Beijing 100084, China 
\\ $^{\dagger}$ Department of Physics, The Ohio State University, Columbus, OH 43210, USA
\\
$^{\dagger \dagger}$ Institute for Advanced Study, Tsinghua University, Beijing 100084, China}


\date{\today}

\begin{abstract}
The advances in cold atom experiments have allowed construction of confining traps in the form of curved surfaces. This opens up the possibility of studying quantum gases in curved manifolds. 
On closed surfaces, many fundamental processes are affected by the local and global properties, i.e. the curvature and the topology of the surface. In this paper, we 
study the problem of potential scattering on a spherical surface and discuss its difference with that on a 2D plane. For bound states with angular momentum $m$, their energies ($E_{m}$) on a sphere are 
related to those on a 2D plane ($-|E_{m,o}|$) as
$E_{m}= - |E_{m, o}|  + E_{R}^{} \left[ \frac{m^2-1}{3} 
 + O\left( \frac{r_o^2}{R^2} \right) \right] $,   where $E_{R}^{} = \hbar^2/(2M R^2)$, and $R$ is the radius of the sphere.  
Due to the finite extent of the manifold, the phase shifts on a sphere at energies $E\sim E_{R}^{}$ differ significantly from 
those on a 2D plane. As energy $E$ approaches zero, the phase shift in the planar case approaches $0$, whereas in the spherical case it reaches a constant that connects the microscopic length scale to the largest length scale $R$. 
 \end{abstract}

\maketitle

In condensed matter physics, one typically deals with systems in Euclidean space. 
In recent years, mesoscopic structures of the form of curved surfaces have also been found in an expanding list of materials, including self-assembled interfaces in lyotopic liquid crystals, carbon nano-fullerences, and bent graphites.
Still, examples of quantum systems in curved surfaces are rare. However, the situation may change due to the
 advances in cold atom experiments. Using dressed adiabatic radio-frequency potential\cite{S1,S2,S3}, one can now construct confining traps in different forms of curved surfaces. 
Loading atoms onto these traps will create quasi two-dimensional quantum gases  in a variety of curved manifolds. 
Alternatively, one can use two immiscible quantum gases
$A$ and $B$. With $A$ forming a large core occupying the center of a harmonic trap, and with $B$ coating the surface of $A$, thereby creating a bubble of quantum gas $B$. 
With the flexibility in controlling densities and interactions in cold atom systems, we now have opportunities to study quantum many-body physics in curved surfaces. 
Due to gravity, atoms in these traps tend to sag to the bottom. Such non-uniform distribution, however, can be eliminated in zero-gravity environment. As a result, the realization of uniform quantum gases in curved surfaces have been considered as possible experiments in Space Stations\cite{Bremen}. 

On a curved surface, some familiar concepts and relations in Euclidean space have to be revised. 
For example, in defining the center of mass of two  particles, one needs to specify the portion of the geodesic connecting them. Moreover, the center of mass motion and relative motion are no longer separable. As a result, the two-body problem in curve space is much more complex.  Curved surface can also produce an effective gauge field for atom pairs with non-zero relative angular momentum\cite{Hui}, in exactly the same way that they produce a superfluid velocity with distributed vorticity 
 in superfluid $^{3}$He-A\cite{He-3} (which is a Berry connection and hence a effective vector potential). 
 
Underlying the two-body problem is the potential scattering of single particle. Here, we focus on this problem and study the effect of curvature. For simplicity, we shall consider the potential scattering on a sphere. 
An immediate question is the relation between the potential scattering on curved surfaces and that on an Euclidean 2D plane. In the latter case, particles can be  either in scattering states or  in bound states. The corresponding  wavefunctions have infinite or finite extent, carrying
positive or negative  energies respectively. Low energy scattering processes are characterized by scattering amplitudes of the form\cite{2Dscattering}
\begin{equation}
f(E) = \frac{2 e^{i\pi/4}}{\sqrt{2\pi k} } \frac{1}{{\rm cot}\delta (k) -i}
\label{f} \end{equation}
where $E>0$ and $\hbar^2 k^2/2M\equiv E$, and $\delta(k)$ is the phase shift\cite{2Dscattering}
\begin{equation}
{\rm cot}\delta (k) = \frac{1}{\pi}{\rm ln}\left(E/E_{o}\right)
\label{cotdelta} \end{equation}
where $E_{o}$ is a positive constant. In addition, the bound state is given by the pole of the scattering amplitude when it is analytically continued to negative energies. 

For curved surfaces, the clear distinction between scattering states and bound states is lost, since the extent of a bound states can extend over the entire manifold as its energy reduces  to the confinement energy $-E_{R}$ of the manifold, where 
\begin{equation}
E_{R}\equiv \hbar^2/(2MR^2).
\end{equation} 
Due to the finite extent of the curved manifold, 
it is useful to separate out two  regimes in length scales, 
\begin{eqnarray}
{\bf I} : &  \,\,\,\,\,  r_{o} < \lambda \ll R,  \,\,\,\,\, 
\label{nearE}\\
{\bf II} :  & \,\,\,\,\,  \lambda \sim R  \,\,\,\,\,
 \label{extended}, 
\end{eqnarray}
where $r_{o}$ is the size of the potential, and $\lambda$ is the wavelength of the extended state or the size of the bound states, 
$\lambda = 2\pi/k = 2\pi \hbar/\sqrt{ 2M |E|}$.  In terms of energy scales, these conditions are 
 \begin{eqnarray}
{\bf I} : &  E_{R}\ll |E| < \hbar^2/(2Mr_{o}^{2}),   
\label{nearE2}\\
{\bf II} :  &  |E| \sim E_{R}
 \label{extended2}, 
 \end{eqnarray}
Regime  $({\bf I})$ will be referred to as the ``near Euclidean" regime. It is the regime where curvature effects are perturbative, and where   connections to Euclidean space can be made. Yet there is an important difference between closed surfaces and the Euclidean plane  in this regime. In the latter case, low energy scattering corresponds to 
$kr_{o}\ll 1$. There is  no upper limit on the wavelength $\lambda = 2\pi/k$. In contrast, $\lambda$ is bounded by the size of the manifold $R$ for closed surfaces. Hence, even though one expects to recover the potential scattering in Euclidean space in regime ${\bf I}$, the recovery will fail as $\lambda$ approaches $R$, where one switches to Regime 
${\bf II}$. Regime ${\bf II}$ will be referred to as the ``curved space" regime, where curvature effects will be significant.  
(See figure 1). The behavior of the scattering phase shift in this regime is unique to closed surfaces. 
\begin{figure}[h]
\centering
\includegraphics[width=7.5cm]{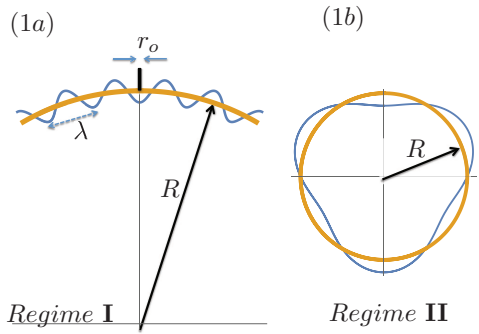} 
\caption{Figure 1a shows the typical wavelength $\lambda$ in the near Euclidean regime, (Regime ${\bf I}$), with $r_{o}< \lambda \ll R$, where $r_{o}$ is the size of the potential and $R$ is the radius of the sphere. Figure 1b shows the typical configuration  in the curved space regime (Regime ${\bf II}$), where $\lambda \sim R$.  }
\label{figure 1}
\end{figure}

Let $\eta_{m}^{}(E)$ and $\delta_{m}^{}(E)$ be the the phase shifts in the angular momentum $m$ channel for potential scattering on a sphere of radius $R$ and on an infinite 2D plane respectively. 
Our key findings are :

\noindent ${\bf (A)}$ In the near Euclidean regime ${\bf I}$, the curvature correction to the s-wave scattering phase shift is 
\begin{equation}
\pi ({\rm cot}\eta_{0}(E) -{\rm cot}\delta_{0}(E)) = \frac{E_{R}^{}}{3E} + O\left(  E_{R} ^2 / E^2 \right),  E\gg E_{R}.
\end{equation}

\noindent ${\bf (B)}$ The curvature correction to the bound state 
 energy in Regime ${\bf I}$ is  

\begin{eqnarray}
E_{m}=- |E_{m, o}| 
+ E_{R}^{} \left[ \frac{m^2-1}{3} 
+ O\left( \frac{r_o^2}{R^2} \right)  \right], 
\end{eqnarray}
where $E_{m}$ and $-|E_{m,o}|$ are the energies of the bound state with angular momentum $m$ in the spherical surface and in the 2D plane created by the same short range potential.

\noindent ${\bf (C)}$ In regime ${\bf II}$, the s-wave scattering phase shift on a sphere $\eta_{0}(E)$ differs significantly from its Euclidean value $\delta_{0}(E)$ over the range $0<E<E_{R}$.  
As $E\rightarrow 0$, ${\rm cot}\delta_{0}(E) \rightarrow -\infty$ whereas 
${\rm cot}\eta_{0}(E)$ remains finite. 

\noindent ${\bf (D)}$ We have also obtained the curvature correction for the phase shifts in angular momentum $m$ channel  up to logarithmic corrections. 

These results are obtained by considering a short range potential $V(\theta)$ in a spherical surface with radius $R$: 
$V(\theta)= 0$ for 
$\theta\geq \theta_{o}$, $\theta_{o}\ll 1$. The analogous potential in a 
2D plane is,  $V(r)=0$ for $r>r_{o}$, with $r_{o}=\theta_{o} R$.
Our results are derived for the limit $\lambda \gg r_{o}$, which is amount to a zero range approximation for the potential. 
Before discussing the spherical case, let us first review some basic features of potential scattering in an infinite 2D plane. 

{\em Short range potential in a 2D plane:} 
Outside the potential $V(r)$, the Schr\"{o}dinger equation is that of a free particle. 
In anticipation to comparing with the spherical case, we artificially introduce a length scale $R$, and rewrite the Schr\"{o}dinger equation in terms of the scaled cylindrical coordinates 
\begin{equation}
 \left[ 
\frac{\partial^2}{\partial \theta^2}+ \frac{1}{\theta}\frac{\partial}{\partial \theta}  + \frac{1}{\theta^2}\frac{\partial^2}{\partial \phi^2}
 \right] \Psi(\theta, \phi) + {\cal E}\Psi(\theta, \phi) =0, \,\,\,\,\, \theta > \theta_{o}, \label{Plane-eq}
\end{equation}
\begin{equation}
\theta \equiv \frac{r}{R}, \,\,\,\, \theta_{o} \equiv \frac{r_{o}}{R}, \,\,\,\, 
{\cal E}=\frac{E}{E_{R}^{}},  \,\,\,\,
 0\leq \theta < \infty,
 \end{equation}
and $(\theta,\phi) = (r/R, \phi)$ are the scaled cylindrical coordinates. We further define 
\begin{equation}
E\equiv \hbar^2 k^2/2M, \,\,\,\, {\rm hence} \,\,\,\,
kR=\sqrt{\cal E}, \,\,\,\,\, 
kr=\sqrt{\cal E} \theta.
\end{equation}
 For a state with angular momentum $m$, its wavefunction outside the potential is   
\begin{equation}
\Psi^{(m)}(\theta,\phi) = A e^{im\phi}\left[ J_{m}(kr) - {\rm tan}\delta_{m}(k) Y_{m}(kr) \right],
\,\,\,  
r>r_{o}
\end{equation}
where $J_{m}$ and $Y_{m}$ are the Bessel functions  of the first and second kind respectively, and $\delta_{m}(k)$ is the phase shift. It is  determined by the logarithmic derivative  at $r_{o}$, 
\begin{equation}
B = r_{o}\left[ \partial_{r} {\rm ln}\Psi^{(m)}(r)\right]_{r=r_{o}}  
= \theta_{o} \left[ \partial_{\theta} {\rm ln}\Psi^{(m)}(\theta)\right]_{\theta=\theta_{o}}
\label{M}\end{equation}
where $B$ is a constant characterized by the property of the potential inside $r_{o}$. 

Let us first consider  s-wave scattering. 
Substituting in Eq.(\ref{M}) the expansion of  $J_{0}$ and $Y_{0}$ at small arguments, the s-wave phase shift is found to be 
\begin{equation}
{\rm cot}\delta_{0}(k) = \frac{2}{\pi}   \left( {\rm ln}(kr_{o}e^{\gamma}/2) - B^{-1} \right)  
= \frac{1}{\pi} {\rm ln}(E/E_{o}), \label{cot}
\end{equation}
where $\gamma =0.577..$ is the Euler constant, and 
$E_{o}$ is related to $r_{o}$ as  
\begin{equation}
E_{o}\equiv \hbar^2 k_{o}^{2}/2M, \,\,\,\,\,\, 
{\rm ln}(k_{o}r_{o}e^{\gamma}/2) \equiv   B^{-1}.
\end{equation}

{\em Short range potential on a spherical surface:} We first define the phase shift for potential scattering on a spherical surface. The Schr\"{o}dinger equation of a particle on a spherical surface outside the potential $V(\theta)$ is 
 \begin{equation}
\left[  
\frac{\partial^2}{\partial \theta^2}+ {\rm cot}\theta \frac{\partial}{\partial \theta}  + \frac{1}{{\rm sin}^2\theta}\frac{\partial^2}{\partial \phi^2}
 \right] \Psi(\theta, \phi)  + {\cal E}\Psi(\theta, \phi) =0
\label{SS} \end{equation}
for  $\theta> \theta_{o}$. The angular momentum eigenstates are of the form  $\Psi^{(m)}(\theta, \phi) = e^{im\phi}\Psi^{(m)}(\theta)$. 
Defining 
\begin{equation}
\mu\equiv {\rm cos}\theta, \,\,\,\, E/E_{R}= {\cal E}\equiv \nu(\nu+1), 
\label{nudef} \end{equation}  
the equation for  $\Psi^{(m)}(\theta)$ becomes
\begin{equation}
\frac{\partial}{\partial\mu} \left[ (1-\mu^2) \frac{\partial}{\partial\mu} \right] \Psi^{(m)} + \left[ \nu(\nu+1) -  \frac{m^2}{1-\mu^2}\right]\Psi^{(m)} = 0, 
\label{Le} \end{equation} 
which is the Legendre equation. 
The general solution is 
\begin{equation} 
\Psi^{(m)}(\theta) = A_m \left( P^{m}_{\nu}(\mu)  + \frac{2}{\pi}{\rm tan}\eta_{m}(\nu) Q^{m}_{\nu}(\mu)\right), \,\,\,\,\,\, 
\theta > \theta_{o}
\end{equation}
where $P^{m}_{\nu}(\mu)$ and $Q^{m}_{\nu}(\mu)$ are the associated Legendre function of the first and second kind respectively, and $\eta_{m}$ is the phase shift. $\eta_{m}$ is a function of $\nu$,  related to the dimensionless energy ${\cal E}$ through Eq.(\ref{nudef}). 
The factor $2/\pi$ was included so as to agree with the planar case in the right limit. The phase shift $\eta_{m}$ is again determined by the boundary of the form of Eq.(\ref{M}).
In terms of dimensionless ``energy" $\nu$ and dimensionless length variable $\theta_{o}$, regime ${\bf I}$ and ${\bf II}$ 
correspond to 
\begin{eqnarray}
{\bf I} : \,\,\,\,\, 1\ll \nu \ll 1/\theta_{o}, \,\,\,\,\, 
 \nu \sim {\cal E}^{1/2}  \label{wc1} \\
{\bf II} : \,\,\,\,\,   |\nu|\sim 1.  \hspace{1.0in}
\label{wc2} \end{eqnarray}

In the absence of external potential, Eq.(\ref{Le}) applies to the entire range $0\leq \theta \leq \pi$.  The Legendre functions become Legendre polynomials $P_{n}(\mu)$, $n=0,1,2, ..$, i.e. the $\nu$'s reduce to non-negative integers. The energies of the s-wave eigenstates are  given by ${\cal E}=n(n+1)$. In the presence of scattering potential, ${\cal E}$ can assumed any value, and $\nu$ is no longer an integer. Rather, it is given by 
\begin{equation}
\nu=(-1 \pm \sqrt{1+4{\cal E}} )/2, 
\label{nu} \end{equation}
which can be real or complex depending on ${\cal E}> -1/4$ or $< -1/4$.
The behavior of the s-wave phase shift $\eta_{0}(\nu)$  in different energy regimes are as follows: 

{\bf (I)}  {\em ${\cal E}> -1/4$ :} In this case, $\nu$ is real. 
To solve for the phase shift $\eta_{0}(\nu)$ in near Euclidean  regime ${\bf I}$, we need to evaluate the boundary condition 
Eq.(\ref{M}) at the small angle $\theta_{o}\ll 1$ using the asymptotic form of the Legendre functions. With the definition 
\begin{equation}
\mu_{o}={\rm cos} \, \theta_{o}, 
\end{equation}
$P^{}_{\nu}(\mu_{o})$ and $Q^{}_{\nu}(\mu_{o})$ assume the forms \cite{nist} \cite{math handbook}
\begin{eqnarray} 
 P^{}_{\nu}(\mu_{o}) = 1- \frac{1}{4}\nu(\nu+1)\theta_{o}^2 +  O(\theta^{4}_{o}), \hspace{0.5in}   \label{as1}\\
\partial_{\mu} P_{\nu}(\mu_{o}) = \nu(\nu+1)/2 + O(\theta^{2}_{o}), \hspace{0.8in}  \label{as2} \\
Q^{}_{\nu}(\mu_{o}) = {\rm ln}\left( \frac{2}{\theta_{o}}\right) - \gamma -  \psi(\nu+1)  + O (\theta_{o}),  \hspace{0.1in}  \label{as3}\\
\partial_{\mu}  Q_{\nu}(\mu_{o}) = \theta^{-2}_{o}  + O( {\rm ln}\theta_{o}), \hspace{1.0in}  \label{as4}
\end{eqnarray}
 where $\psi(\nu)$ is diGamma function. Substituting these asymptotic forms into 
Eq.(\ref{M}), one obtains
\begin{equation}
{\rm cot}\eta_{0}(\nu) =  (- 2/\pi)\left( Q_{\nu}(\mu_{o}) +  B^{-1} \right).
\label{coteta} \end{equation}
To write out $ Q_{\nu}(\mu_{o})$ explicitly in regime ${\bf I}$, 
where $\nu\gg 1$,  we explore the asymptotic form the diGamma function in Eq.(\ref{as3}) for large $\nu\gg 1$ 
\begin{equation}
\psi(\nu +1) = {\rm ln}\nu + \frac{1}{2\nu} - \frac{1}{12 \nu^2} + O(\nu^{-4}).  \label{as5}
\end{equation}
With the expansion  $\nu = \sqrt{\cal E} - 1/2 + 1/(8 \sqrt{\cal E}) +O({\cal E} ^{ -3/2}) $ , we have  
\begin{equation}
 Q_{\nu}(\mu_{o})  = {\rm ln} \left( \frac{2}{\theta_{o}\nu e^{\gamma}}\right)  - 
\frac{1}{2\nu} + \frac{1}{12\nu^2} + O({\cal E} ^{ -2 }) .
\label{Q} 
\end{equation}
Substituting Eq.(\ref{Q}) into 
Eq.(\ref{coteta}), and replacing $\nu$ by ${\cal E}$, we have  
\begin{equation}
\pi {\rm cot} \eta_{0}(E) = {\rm ln}\left(E/E_{o} \right) + \frac{1}{3 {\cal E}}  + O({\cal E} ^{ -2 }) .
\label{cotd} 
\end{equation}
where $E_{o}$ is the quantity appearing in Eq.(\ref{cot}).
The first term in Eq.(\ref{cotd}) is the phase shift in 2D Euclidean space, Eq.(\ref{cot}), the second is the curvature correction. Eq.(\ref{cotd}) is the result ${\bf (A)}$ mentioned in the introduction.

\begin{figure}[h]
\centering
\includegraphics[width=7.5cm]{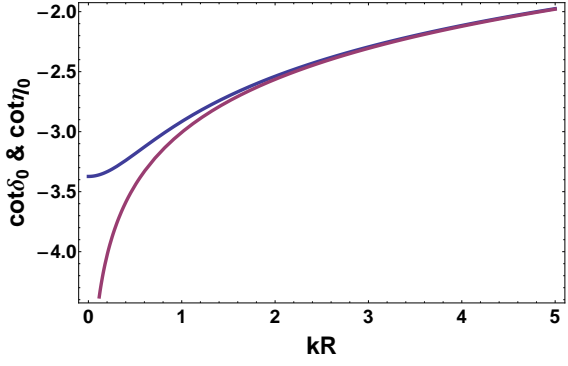} 
\caption{ ${\rm cot}\delta_{0}(E)$ and ${\rm cot}\eta_{0}(E)$ versus dimensionless parameter $kR\equiv \sqrt{E/E_{R}}$:  The blue and red curves are  the values of ${\rm cot}\delta_{0}(E)$  for  2D plane and of ${\rm cot}\eta_{0}(E)$ for 2D spherical surface respectively.  The near-Euclidean regime for energy is  $1\ll kR\ll 1/\theta_{o}$. In our calculation,  we have used $\theta_{o}=0.01$. We see in this plot that the difference between the two phase shifts is small in the near Euclidean range  $(kR> 3)$, but becomes increasingly large as energy is decreased. This increase in  difference is due to the presence of the extended states  that cover the entire spherical surface which have no analog in the 2D plane.   }
\label{fig:shift-compare}
\end{figure}

In the curved-space regime {\bf II}, the eigenstates extend over the entire sphere. The 
asymptotic forms Eq.(\ref{as5})-(\ref{cotd}) no longer apply. The phase shift has to be obtained by solving Eq.(\ref{M}) numerically.  In figure \ref{fig:shift-compare}, we show that the phase shift $\delta(k)$ of both the spherical and planar geometry for a hard core potential, as calculated from Eq.(\ref{M}). In this case, 
$B^{-1}=0$. 

The major difference between $\eta_{0}(E)$ and 
$\delta_{0}(E)$ (for all boundary condition $B^{-1}$ ) occurs at $E\rightarrow 0$. In a 2D plane, Eq.(\ref{cotdelta}) shows 
  that $\lim_{E=0} \delta_0 = 0^{-} $ for any finite $r_{o}$. To find the phase shift $\eta_{0}$ at $E=0$, we note that the diGamma function in Eq.(\ref{as3}) has the limit 
$\psi(\nu+1) = - \gamma +O(\nu) $ as $\nu\rightarrow 0$. Eq.(\ref{coteta}) can then be reduced to 
\begin{equation}
{\rm cot}\eta_{0}(E\rightarrow 0) \simeq
-\frac{2}{\pi} \left[ {\rm ln}\left( \frac{2}{\theta}_{o}\right) + B^{-1} \right]. 
\end{equation}
The phase shift $\eta_{0}(E\rightarrow 0)$ is a constant that depends on $1/\theta_{o}$, i.e., the ratio $r_{o}/R$, 
whereas in a 2D plane.
This is the result ${\bf (C)}$ mentioned in the introduction.

{\bf (II)}  {\em ${\cal E}< -1/4$ : } In this case, $\nu$ is complex. Eq.(\ref{nu}) implies $\nu = -1/2 + i\tau$, with $\tau\equiv |1/4+{\cal E}|^{1/2}$.  In particular, the deep bound states (with ${\cal E}\ll -1$) correspond to $\tau = \sqrt{|{\cal E}|}+O(|{\cal E}|^{-1/2})$ to the 0-th order. For complex $\nu$, $P_{\nu}(\mu)$ is real but  $Q_{\nu}(\mu)$ is complex. Moreover, both functions diverge at the south pole, $\mu\rightarrow -1^{+}$.  Thus, to construct the eigenstates of Eq.(\ref{Le}),  and to evaluate the boundary condition Eq.(\ref{M}), we need to examine the asymptotic forms of these functions at the south pole  so as to construct real and finite solutions of Eq.(\ref{Le}) for $\theta >\theta_{o}$.  The asymptotic forms of these functions near the south pole (i.e.,$ \mu \rightarrow -1^{+} $) are 
\begin{eqnarray}
P_{\nu}(\mu ) =  - {\rm ln}(1+\mu)/2  +O(1)   \hspace{0.6in}
\\
{\rm Re} Q_{\nu}(\mu) = \frac{\pi}{ 2 {\rm cosh}(\tau \pi) } + O(\mu+1)   \hspace{0.5in} \\
{\rm Im} Q_{\nu}(\mu ) = - \frac{\pi}{ 2} {\rm tanh}(\tau \pi)   P_{\nu}(\mu )  +O(\mu+1).   \hspace{0.2in}
\end{eqnarray}
Thus, the only real solution finite for all $\theta >\theta_{o}$  is $\Psi(\theta) = Re[Q_{\nu}(\mu)]$.  
( The real functions $P_{-\frac{1}{2}+i\tau}(\mu)$ and  $Re[Q_{-\frac{1}{2}+i\tau}(\mu)]$ are known to be the conical functions\cite{nist}. )
For deep bound states, ${\cal E}\ll -1/4$, or $\tau \gg 1$, 
one expects they decay exponentially with $\theta$. 
This can be seen in the behavior of $Re[Q_{-\frac{1}{2}+i\tau}(\mu)]$. In the region $\theta <\pi/2$, it is 
\begin{equation}
{\rm Re}Q_{\nu}(\mu) = \left( \frac{\theta}{{\rm sin}\theta}\right)^{1/2}
K_{0}(\tau \theta) \left( 1 + O(1/\tau) \right) 
\end{equation}
where $K_{0}$ is the Bessel function of imaginary argument, 
and $K_{0}(\tau \theta ) \propto  e^{-\tau \theta}$   for $\tau\theta \gg 1$. 
The bound state energy can be found by numerically solving Eq.(\ref{M}) for given $B$, with $\Psi(\theta) = Re[Q_{\nu}(\mu)]$. Surely, for bound states with bound state energy $|E| \gg  E_{R}$, the curvature effects are only perturbative. 

{\em Finding bound states from analytic continuation of the phase shift:} For potential scattering in Euclidean space, it is known that a bound state will show up as a pole in the scattering amplitude when it is continued analytically from positive to negative energy. With the expression of scattering amplitude given in Eq.(\ref{f}), the bound state energy is given by 
\begin{equation}
{\rm cot}\delta_{0}(-|E|) -i =0. 
\label{bdd}  \end{equation}

Using the expression of phase shift with curvature correction in  near Euclidean regime ${\bf I}$, 
Eq.(\ref{cotd}), this condition becomes 
\begin{equation}
0=\frac{1}{\pi} \left(   {\rm ln}\left(-|E|/E_{o} \right) + \frac{1}{3(-|{\cal E}|)} \right)  -i 
\end{equation} 
which gives the bound state energy  
\begin{equation}
E= -E_{o}- E_{R}/3 +  O(E_{R}^2/E_{o} ^2) E_{o}  <0 \label{E_b}
\end{equation}
Note that Eq.(\ref{E_b}) is valid when $E_{o}\gg E_{R}$, which is 
guaranteed in the near Euclidean regime ${\bf I}$.
The result Eq.(\ref{E_b}) has also been confirmed by the numerical solution of Eq.(\ref{M}). 

{\em A simple derivation of the curvature corrections for the energies of deep bound states :} As far as curvature correction to bound state energy is concerned, it can be obtained in a  simpler way. Expanding Eq.(\ref{SS}) around the north pole, 
($\theta=0$), the Schr\"{o}dinger equation Eq.(\ref{SS}) in the region $\theta>\theta_{o}$ can be expanded in powers of curvature  as 
\begin{equation}
\left( - \nabla^{2}  + U(\theta, \phi) \right)\Psi(\theta, \phi)  =  {\cal E}\Psi(\theta, \phi),  
\label{S4} 
\end{equation}
\begin{equation}
U(\theta, \phi) =  \frac{1}{3} \left( \theta \frac{\partial}{\partial \theta} - \frac{\partial^2}{ \partial \phi^2 } \right)
\label{U} \end{equation}
where $\nabla^2$ is the Laplacian operator in the 2D plane as shown in Eq.(\ref{Plane-eq}). If $\Phi_{m,o}$ is  a bound state in the 2D plane with angular momentum $m$ and binding energy $-|{\cal E}_{m,o}|\equiv - |E_{(m),o}|/E_{R}$, then the bound state energy ${\cal E}$ of Eq.(\ref{S4}) can be obtained by doing the first order perturbation on $U$,   
\begin{eqnarray}
{\cal E} & = &  -|{\cal E}_{m,o}| + \int^{2\pi}_{0}{\rm d}\phi\int^{\pi}_{0} \theta {\rm d}\theta \,\, 
\Phi_{m,o}^{\ast} U(\theta, \phi) \Phi_{m,o} \\
& = & -|{\cal E}_{m,o}| + \frac{ m^2 -1}{3} 
\label{calE}
\end{eqnarray}
This is the result ${\bf (B)}$ mentioned in the introduction.

{\em Higher angular momentum states:}  For states with non-zero angular momentum $m$, one can obtain its phase shift $\eta_{m}(E)$ from Eq.(\ref{M}) using the asymptotic form of the associated Legendre functions at short distances. The latter ones with $m>0$ are \cite{nist} \cite{math handbook}
\begin{eqnarray} 
P^{m}_{\nu}(\mu_{o})=\frac{\Gamma(\nu+m+1)}{\Gamma(\nu-m+1)}\frac{(-\theta_{o})^m}{m! 2^m} +O(\theta_{o}^{m+2}) \label{m1} \\
\partial_{\theta}  P^{m}_{\nu}(\mu_{o}) =  \frac{m}{\theta_{o}}  P^{m}_{\nu}(\mu_{o}) + O(\theta^{m+1}_{o})  \hspace{1.5 cm} \\
Q^{m}_{\nu}(\mu_{o}) = \frac{1}{2} (-1)^m \Gamma (m) \left(\frac{2}{\theta_{o}}\right)^m + O(\theta_{o}^{2-m}) \\
\partial_{\theta} Q^{m}_{\nu}(\mu_{o}) = \frac{-m}{\theta_{o}}  Q^{m}_{\nu}(\mu_{o}) + O(\theta^{-m+1}_{o}). \hspace{1 cm} 
\label{m4} 
\end{eqnarray}
For those with minus angular momentum, we have used the relation  
\begin{equation}
P^{-m}_{\nu}=(-1)^m\frac{\Gamma(\nu-m+1)}{\Gamma(\nu+m+1)}P^{m}_{\nu},
\end{equation}
and the same relation between  $Q^{-m}_{\nu}$ and  $Q^{m}_{\nu}$. 
To obtain the phase shift in near Euclidean regime ${\bf I}$,  where $\nu \gg 1$, we need to apply the asymptotic expansions Eq.(\ref{m1}) to (\ref{m4}) to large $\nu$. In this limit, we have  
 $\frac{\Gamma(\nu-m+1)}{\Gamma(\nu+m+1)} = \mathcal{E}^{m} \left(1+ \frac{m-m^3}{3\mathcal{E}}+O( \mathcal{E}^{-2}) \right)$, where $\mathcal{E} \gg 1 $, 
we then find from Eq.(\ref{M})   
\begin{eqnarray}
\pi \cot \eta_m &=&\frac{m\Gamma(m)^2 } { (kr_{o}/2)^{2m}} \left[1+ \frac{m-m^3}{3k^2R^2}\right]^{-1} \left[ \frac{m+B}{m-B} \right]  \nonumber  \\
   &  & +O\left(  (kr_{o})^{2-2m} \right)  \label{m} 
\end{eqnarray}
where  $kr_{o}=\sqrt{{\cal E}}\theta_{o} \ll 1$.   
The hard core potential correspond to the case  $B^{-1}=0$.

In the 2D planar case, similar calculation shows 
\begin{equation}
\pi \cot \delta_m=\frac{m\Gamma(m)^2 } {(kr_{o}/2)^{2m}} \left[ \frac{m+B}{m-B}\right] + O\left(   (kr_{o})^{2-2m} \right) \,\, . 
\end{equation}
This can be derived straightforwardly from Eq.(\ref{M}) using the Bessel Function asymptotic behavior at short distance $J_{m}(x) \simeq [\Gamma(m+1)]^{-1}(x/2)^m$ and $Y_{m}(x) \simeq -\Gamma(m)(x/2)^{-m}/\pi$.  The $\frac{m-m^3}{3k^2R^2}$ term in Eq.(\ref{m}) is the curvature correction mentioned in ${\bf (D)}$ in the introduction. 
We have not been able to obtain the higher order corrections for $\delta_{m}$ which consist of log-corrections. 
Such corrections are needed to obtain the bound state energies  by analytic continuation of the scattering amplitudes. 

In summary, we have studied the potential scattering on a spherical surface and have obtained the curvature corrections to the phase shifts and bound state energy, as summarized in statements ${\bf (A)}$ to ${\bf (D)}$ earlier.  These exact results can serve as a calibration of effective field theory approaches on curved surfaces. The curvature corrections to the bound state energy also make an interesting conceptual point. If in a "universe" where one was able to determine the energy shift of a bound state at different locations, one might be able to map out the geometry of this ``universe" through precision measurements.

{\em Acknowledgement:} this work is supported by
the National Key R$\&$D Program of China under Grant No. 2017YFA0304900, 
the NSFC Grant Nos. 11204153 and 11674192; and 
the NSF Grant DMR-0907366, the MURI Grant FP054294-D, and the NASA Grant on Fundamental Physics 1541824.

\end{document}